%% file: ANA-HIGP-2024-04-PAPER.tex
\documentclass[PAPER, atlasdraft=false, USenglish, cernpreprint, texmf, orcidlogo]{atlasdoc}
\usepackage[subfigure]{atlaspackage}
\usepackage{atlasbiblatex}

\usepackage{atlasphysics}

\usepackage{multirow}

\graphicspath{{logos/}{figures/}}

\AtlasTitle{Search for Higgs bosons produced in association with a high-energy photon via vector-boson fusion and decaying to a pair of \(b\)-quarks in the ATLAS detector}

\AtlasAbstract{%
A search for Standard Model Higgs bosons produced in association with a high-energy photon and decaying to $b\bar{b}$ is performed using 133 fb\(^{-1}\) of \(\sqrt{s}=13\)~TeV \(pp\) collision data collected with the ATLAS detector at the Large Hadron Collider at CERN.
The photon requirement reduces the multijet background, and the \(H \rightarrow b\bar b\) decay is the dominant decay mode.
Event selection requirements target events produced by vector-boson fusion, the dominant production mode in this channel.
Several improvements enhance the search sensitivity compared to previous measurements.
These improvements include better background modelling and characterization, the use of a neural-network classifier, and an updated signal extraction strategy adopting a direct binned-likelihood fit to the classifier output.
With these improvements, the Higgs boson signal strength is measured to be \(0.2 \pm 0.7\) relative to the Standard Model prediction.
This corresponds to an observed significance of \(0.3\) standard deviations, compared to an expectation of \(1.5\) standard deviations assuming the Standard Model.
}

\AtlasRefCode{HIGP-2024-04}

\PreprintIdNumber{CERN-EP-2025-110}

\AtlasDate{21 August 2026}

\arXivId{2509.14005}

\HepDataRecord{169580}

\AtlasJournal{Phys.\ Lett.\ B.}
\AtlasJournalRef{\PLB 877 (2026) 140279}
\AtlasDOI{10.1016/j.physletb.2026.140279}

\hypersetup{pdftitle={ATLAS document},pdfauthor={The ATLAS Collaboration}}

\begin{document}

\maketitle

\tableofcontents

\section{Introduction}
\label{sec:intro}

\begin{figure}[t]
\centering
\subfigure[]{
\includegraphics[width=0.4\textwidth]{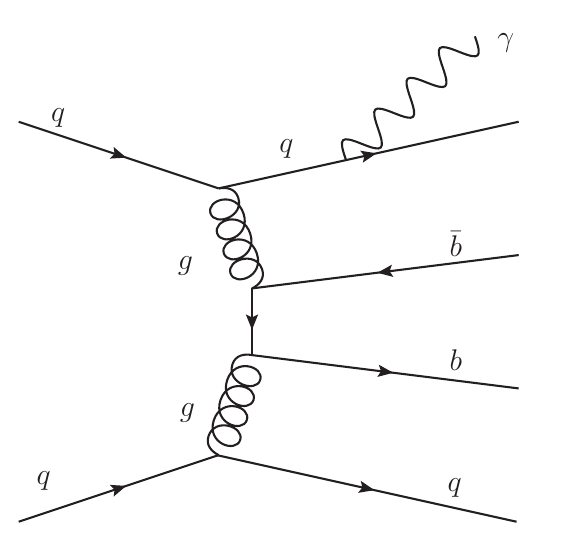}
\label{fig:Diagrams_a}
}
\subfigure[]{
\includegraphics[width=0.4\textwidth]{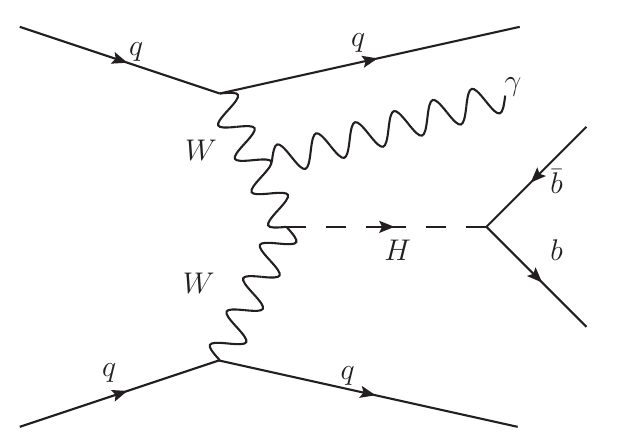}
\label{fig:Diagrams_b}
}
\caption{Representative Feynman diagrams for (a) non-resonant multijet background and (b) \(H\gamma jj\) signal.
The non-resonant multijet background can also have a \(qg\) initial state, and the photon can be radiated from an initial-state or final-state quark.}
\end{figure}

The scalar Higgs boson, discovered in 2012~\cite{HIGG-2012-27, CMS-HIG-12-028}, has been studied extensively in all the expected main production modes and in many of its decay channels.
The Higgs boson decays most frequently to a pair of \(b\)-quarks, making this a powerful decay channel to study.
However, this decay is challenging because of a significant non-resonant QCD multijet background, which complicates the search.
By requiring a final state with an associated high-energy photon, a destructive interference effect, arising from diagrams with initial-state and final-state radiation of the photon, reduces this non-resonant background, making this an interesting final-state signature to target~\cite{Gabrielli:2007wf}.
An example Feynman diagram for this dominant non-resonant background is shown in Figure \ref{fig:Diagrams_a}.
Conveniently, the photon also provides a clean signature on which to trigger.
Due to the increased number of electroweak vertices in vector-boson fusion (VBF) Higgs boson production, this production mode surpasses gluon--gluon fusion (\(ggH\)) and becomes the dominant production mechanism when an associated photon is required.
The leading-order Feynman diagram for this mode is shown in Figure \ref{fig:Diagrams_b}.
By requiring a high dijet invariant mass, the signal phase space becomes purer in VBF Higgs production, with vector-boson associated Higgs production (\(VH, V=W,Z\)) contributing less than 5\% of the signal.
Additionally, a destructive interference effect, similar to that reducing the non-resonant QCD multijet background, suppresses \(ZZ\)-fusion and makes this a unique way to study \(WW\)-fusion Higgs production and the \(HWW\) interaction more closely.

Searches for VBF Higgs production with decay into \(b\)-quarks without the associated photon requirement have been performed by both ATLAS and CMS.
In the all-hadronic channel, ATLAS measured a signal strength, or the ratio between the observed signal yield and that predicted by the Standard Model, of \(0.95^{+0.38}_{-0.36}\)~\cite{HIGG-2019-04}.
In a boosted (high transverse momentum \pT) VBF \(H(\rightarrow b\bar{b})\) search, CMS measured a signal strength of \(4.9^{+1.9}_{-1.6}\)~\cite{CMS-HIG-21-020}.
This letter presents an analysis of \(\SI{133}{\ifb}\) of \(\sqrt{s}=\SI{13}{\TeV}\) proton--proton (\(pp\)) collision data collected with the ATLAS detector in search of the VBF \(H(b\bar{b}) + \gamma\) signature.
The VBF Higgs boson signal with the requirement of an associated photon with transverse momentum of at least \(\SI{19}{\GeV}\) and two additional jets (VBF H$\gamma$jj) with transverse momentum of at least \(\SI{15}{\GeV}\) has a cross-section of \(55.6^{+0.6}_{-0.4}\,\)fb predicted at next-to-leading order (NLO).
A previous search for this signature using the same dataset observed a signal strength of \(\mu_H = 1.3\pm1.0\), with an expected significance of \(1.0\) standard deviations (\(1.0\sigma\))~\cite{HIGG-2020-14}.
Since the previous effort, the multivariate analysis technique has been markedly improved, replacing the previously used boosted decision tree (BDT) classifier with a densely connected neural network (NN) and including the discriminating power of kinematic input variables related to the di-\(b\)-jet invariant mass~\cite{DENBY1988429}.
The data-fitting approach is also quite different from that adopted previously; it now extracts the signal strength directly from the classifier's output score distribution and uses the di-\(b\)-jet invariant mass (\(m_{bb}\)) to define regions of purity, rather than fitting the \(m_{bb}\) distribution.
These advances, combined with much larger simulated event samples for the main backgrounds, reduce the signal strength uncertainty and provide improved signal sensitivity.

Details of the analysis procedure are provided in the following sections.
Attributes of the experimental detector are given in Section~\ref{sec:detector}.
Particulars of the simulated data are presented in Section~\ref{sec:mcsamples}.
Characteristics of data collection and event reconstruction and selection are delineated in Section~\ref{sec:selection}.
The techniques for separating this rare signal from common backgrounds are described in Section~\ref{sec:mva}, followed in Section~\ref{sec:modeling} with the specifics of ensuring that such backgrounds suitably model the data.
Systematic uncertainties related to theory modelling and experimental detector effects are catalogued in Section~\ref{sec:systematics}, and the results are presented in Section~\ref{sec:result}.

\section{ATLAS detector}
\label{sec:detector}
\newcommand{\AtlasCoordFootnote}{%
ATLAS uses a right-handed coordinate system with its origin at the nominal interaction point (IP) in the centre of the detector and the \(z\)-axis along the beam pipe.
The \(x\)-axis points from the IP to the centre of the LHC ring, and the \(y\)-axis points upwards.
Polar coordinates \((r,\phi)\) are used in the transverse plane, \(\phi\) being the azimuthal angle around the \(z\)-axis.
Tranverse momentum (\pT) and energy (\ET) are the components of momentum and energy in this transverse plane, perpendicular to the beamline.
The pseudorapidity is defined in terms of the polar angle \(\theta\) as \(\eta = -\ln \tan(\theta/2)\) and is equal to the rapidity \(y = \frac{1}{2} \ln \left( \left(E + p_z \right)/\left(E - p_z \right) \right)\) in the relativistic limit.
Angular distance is measured in units of \(\Delta R \equiv \sqrt{(\Delta y)^{2} + (\Delta\phi)^{2}}\).}

The ATLAS experiment~\cite{PERF-2007-01} at the Large Hadron Collider (LHC) is a multipurpose particle detector with a forward--backward symmetric cylindrical geometry and nearly \(4\pi\) coverage in solid angle.\footnote{\AtlasCoordFootnote}
It consists of an inner tracking detector surrounded by a thin superconducting solenoid providing a \qty{2}{\tesla} axial magnetic field, electromagnetic and hadronic calorimeters, and a muon spectrometer.
The inner tracking detector covers the pseudorapidity range \(|\eta| < 2.5\).
It consists of silicon pixel, silicon microstrip, and transition radiation tracking detectors.
Lead/liquid-argon (LAr) sampling calorimeters provide electromagnetic (EM) energy measurements with high granularity within the region \(|\eta|< 3.2\).
A steel/scintillator-tile hadronic calorimeter covers the central pseudorapidity range (\(|\eta| < 1.7\)).
The endcap and forward areas are instrumented with copper/LAr and tungsten/LAr calorimeters for EM and hadronic energy measurements in the region \(3.2 \leq |\eta| \leq 4.9\).
The muon spectrometer surrounds the calorimeters and is based on three large superconducting air-core toroidal magnets with eight coils each.
The field integral of the toroids ranges between \num{2.0} and \qty{6.0}{\tesla\metre} across most of the spectrometer.
The muon spectrometer includes a system of precision tracking chambers up to \(|\eta| = 2.7\) and fast detectors for triggering up to \(|\eta| = 2.4\).
The luminosity is measured mainly by the LUCID-2 detector, which is located close to the beampipe.
A two-level trigger system is used to select events~\cite{TRIG-2016-01}.
The first-level trigger (L1) is implemented in hardware and uses a subset of the detector information to accept events at a rate below \qty{100}{\kHz}.
This is followed by a software-based trigger (HLT) that reduces the accepted event rate to around \qty{1.25}{\kHz} on average depending on the data-taking conditions.
A software suite~\cite{SOFT-2022-02} is used in data simulation, in the reconstruction
and analysis of real and simulated data, in detector operations, and in the trigger and data acquisition systems of the experiment.

\section{Monte Carlo event simulation samples}
\label{sec:mcsamples}

Fully simulated $H(\rightarrow b\bar{b}) + \gamma$ Monte Carlo (MC) signal events were generated at NLO in QCD with \MGNLO~\cite{Alwall:2014hca} and showered using \HERWIG[7]~\cite{Bahr:2008pv,Bellm:2015jjp}, with the PDF4LHC15 parton distribution function (PDF) set~\cite{Butterworth:2015oua}.
This signal sample consists of Higgs bosons produced through VBF or $VH$ processes.
The set of tuned parameters is assigned the \HERWIG7 default values of \texttt{H7-PS-MMHT2014LO} and \texttt{H7-UE-MMHT}.
Contributions from Higgs processes that are rarer in the selected phase space were considered separately but are treated as background.
To describe the $H+\gamma$ component of other Higgs production processes, photon-filtered samples are derived from inclusive productions samples.
The $ggH$ sample used \POWHEG~\cite{Nason:2004rx,Frixione:2007vw,Alioli:2010xd} interfaced with \PYTHIA[8]~\cite{Sjostrand:2014zea}.
Higgs production in association with a $t\bar{t}$ pair ($ttH$) is described with MC simulated events also generated using \POWHEG interfaced with  \PYTHIA[8] and the NNPDF3.0 PDF set~\cite{Ball:2014uwa}.

For the main non-resonant backgrounds, $b\bar{b}\gamma jj$ and $c\bar{c}\gamma jj$, large samples of fully-simulated Monte Carlo background events were generated at leading order (LO) specifically for this analysis, using \MGNLO for the matrix element calculation and \PYTHIA[8] for the showering to final-state particles.
Both of these samples used the PDF set PDF4LHC21~\cite{Ball:2022oua}, and a restriction requiring dijet invariant mass above $\SI{500}{GeV}$ was applied to improve event selection efficiency in the `offline' phase space discussed in Section \ref{sec:selection}.
This truth-level dijet invariant mass requirement selects about 37\% of all non-resonant QCD background events, which are then passed to the detailed detector simulation, without biasing the results.
No events with truth-level invariant mass below $\SI{500}{GeV}$ have reconstructed-level invariant mass passing the offline requirement (>800 GeV) described in the next section.
The final non-resonant background considered, $t\bar{t}$ production, is modelled by samples with all-hadronic or either partially or fully leptonic final states that were generated at NLO using \POWHEG and showered using \PYTHIA[8].
These samples were normalized to the next-to-next-to-leading order (NNLO) cross-section predictions~\cite{Czakon:2011xx}.
Electroweak $Z\gamma jj$ events were modelled specifically for this analysis at NLO using the same parameter values and PDF set as the signal sample.
QCD $Z\gamma jj$ events were generated at LO using \MGNLO for the hard-scattering and \PYTHIA[8] for showering.

Certain simulation configurations are common to all samples.
The decays of bottom and charm hadrons were performed by \textsc{EvtGen}~\cite{Lange:2001uf}.
Fully simulated minimum-bias events were generated using the \PYTHIA[8] generator~\cite{Sjostrand:2007gs} with the \textsc{NNPDF2.3lo} PDF set and the A3 set of tuned parameters~\cite{ATL-PHYS-PUB-2016-017}, and then overlaid on each hard-scatter interaction, in accord with the luminosity profile of the recorded data, to model pileup contributions from the same bunch crossing and neighbouring bunch crossings.
All simulated samples include a full simulation of the ATLAS detector and its response~\cite{SOFT-2010-01} based on \textsc{Geant4}~\cite{Agostinelli:2002hh}.
Generator choices, the orders of their calculations, and PDF sets used for the considered MC samples are summarized in Table \ref{tab:mcxsec}.

\begin{table}[hbpt]
\caption{Generators, orders of calculation, and PDF sets used for the MC signal and background samples.
The short name of \textsc{aMC@NLO} is used for \MGNLO.}
\label{tab:mcxsec}
\centering
\resizebox{\columnwidth}{!}{\begin{tabular}{l l l l }
\hline
\hline
Sample    & Generator  & Calculation order & PDF set  \\
\hline
$H(\rightarrow b\bar{b})+\gamma$ & \textsc{aMC@NLO}+\HERWIG7 & NLO  & PDF4LHC15   \\ %
$ggH+\gamma$ & \POWHEG NNLOPS+\PYTHIA8  & NNLO & PDF4LHC15 \\
$ttH$ (all-hadronic) & \POWHEG+\PYTHIA8 & NNLO & NNPDF3.0 \\
$ttH$ (semileptonic) & \POWHEG+\PYTHIA8 & NNLO & NNPDF3.0 \\
\hline
$Z(\rightarrow b\bar{b})\gamma jj$ (EWK)   &  \textsc{aMC@NLO}+\HERWIG7  & NLO & PDF4LHC15     \\
$Z(\rightarrow b\bar{b})\gamma jj$ (QCD)   & \MADGRAPH5+\PYTHIA8  & LO  & PDF4LHC15    \\
$b\bar{b}\gamma jj$ (non-resonant) & \MADGRAPH5+\PYTHIA8  & LO & PDF4LHC21   \\
$c\bar{c}\gamma jj$ (non-resonant) & \MADGRAPH5+\PYTHIA8  & LO & PDF4LHC21     \\
$t\bar{t}$ (non-all-hadronic) & \POWHEG+\PYTHIA8 & NLO & NNPDF3.0 \\
$t\bar{t}$ (all-hadronic) & \POWHEG+\PYTHIA8 & NLO & NNPDF3.0 \\
\hline
\hline
\end{tabular}
}
\end{table}

\section{Object and event selection}
\label{sec:selection}

The target signature of this analysis consists of two jets passing the requirements of a flavour-tagging algorithm (`b-tagged jets') from the $b$-quarks from the Higgs boson decay, of two additional jets (`VBF-candidate jets') with large invariant mass originating from the quarks that produced the hard scattering after the emission of the $W$ or $Z$ bosons that fused into a Higgs boson, and of a high-energy photon candidate.
Isolated leptons are vetoed to maintain orthogonality with other ATLAS Higgs boson analyses.
The selection requirements for these objects, which balance high signal acceptance with high background rejection, are summarized in Table \ref{tab:eventselection}.

This analysis uses the full Run-2 dataset recorded from 13 TeV \(pp\) collisions by the ATLAS detector from 2015 to 2018.
Several triggers were chosen and combined to target the VBF phase space and maximize the acceptance.
One motivation for requiring an associated photon in this Higgs boson search is the clean signature the photon provides for triggering, so each trigger chosen in the HLT requires at least one central photon with $p_{\mathrm{T}} > \SI{25}{GeV}$~\cite{TRIG-2018-05}.
These triggers also require at least four jets with $p_{\mathrm{T}} > \SI{35}{GeV}$.
Since the VBF topology tends to produce final-state jets with large angular separation and high invariant mass, the triggers required the largest dijet mass to be at least $\SI{700}{GeV}$.
Starting from late 2016, these triggers also required at least one jet to be $b$-tagged at the 77\% efficiency working point of the MV2c20 or MV2c10 algorithm~\cite{TRIG-2018-08}.
Because of changes in trigger availability and data-quality requirements~\cite{DAPR-2018-01}, particularly in the early Run-2 period, this analysis uses a total integrated luminosity of $\SI{133}{\ifb}$, slightly less than the $\SI{140}{\ifb}$ used in other ATLAS full Run-2 analyses.

Interaction vertices from $pp$ collisions are reconstructed from at least two tracks with $\pT > \SI{500}{MeV}$ that are consistent with originating from the beam collision region.
If more than one interaction vertex is found in the event, the vertex with the largest sum of track squared $\pT$ is selected as the primary vertex~\cite{ATL-PHYS-PUB-2015-026}.
Due to trigger requirements, each of the four jets selected offline are required to have $p_{\mathrm{T}} > \SI{40}{GeV}$.
Jets are reconstructed using the \antikt algorithm~\cite{Cacciari:2008gp,Fastjet} with a radius parameter of \(R=0.4\), using as inputs `particle flow' objects determined by matching the energy deposits in the calorimeter with tracks in the inner detector.
This choice of particle flow objects leads to better resolution and pileup rejection~\cite{PERF-2015-09} compared to Ref.~\cite{HIGG-2020-14}, for which the jet clustering inputs were topological clusters of neighbouring cells in the calorimeters with significant energy deposits.
To enhance pileup rejection, jets in the central region must be tightly matched to the event's primary vertex according to a likelihood-based discriminant called the jet-vertex tagger (JVT), while the similar forward jet-vertex tagger accepts a looser matching for jets in the forward region~\cite{PERF-2014-03,ATL-PHYS-PUB-2019-026}.

This analysis replaces the MV2c10 offline $b$-tagging algorithm, used in the previous analysis~\cite{HIGG-2020-14}, with the more advanced DL1r algorithm, which employs a deep neural network for jet flavour classification~\cite{FTAG-2019-07}.
The pseudorapidity of $b$-tagged jets must fall within the acceptance of the inner detector ($|\eta|<2.5$), on which flavour-tagging heavily relies.
Jets are $b$-tagged if they pass a fixed-cut working point of the DL1r tagging discriminant, which was calibrated using $t\bar{t}$ events to give 77\% efficiency for true $b$-jets with $\pT \geq \SI{20}{\GeV}$.
The 77\% efficiency working point provides rejection factors of 6 and 134 for charm and light-flavour jets, respectively.
Per-jet scale factors are applied in MC events to account for efficiency differences between data and simulation~\cite{FTAG-2018-01,FTAG-2020-08,FTAG-2019-02}.
In the simulation, $b$-tagging is fully emulated (`direct tagging') for all jets for signal events and for events from background processes other than $b\bar{b}\gamma jj$ and $c\bar{c}\gamma jj$.
Simulated events that do not have at least two $b$-tagged jets are rejected outright.
In contrast, the two main backgrounds, $b\bar{b}\gamma jj$ and $c\bar{c}\gamma jj$, undergo `truth-tagging' to maximize the number of simulated events used in the subsequent steps of the analysis.
Each event is assigned a weight based on the probability of each jet in the event to pass or fail the $b$-tagging requirement based on their kinematic properties and flavour of the originating quark.
These tagging approaches yield kinematic distributions that differ negligibly in shape, but noticeably in normalization due to the online $b$-tagging trigger requirement which forces a direct-tagging scheme for at least one of the $b$-jet candidates~\cite{ATL-PHYS-PUB-2022-041}.
This normalization effect is unimportant because the normalization of these two backgrounds combined is determined by a fit to data, as described in Section \ref{sec:modeling}.
To compensate for $b$-jet energy losses due to semileptonic decays and small out-of-cone energy deposits, jet energy corrections are applied, including corrections for muons or undetected neutrinos from such decays~\cite{JETM-2022-01}.
The two $b$-tagged jets with highest transverse momentum are selected as Higgs boson decay-product candidates.
The remaining jets with \pT greater than 40 \GeV are combined pairwise, and the VBF-candidate jets are selected as the pair with the highest invariant mass, requiring that this mass exceeds $\SI{800}{GeV}$, where the trigger efficiency reaches a plateau.

The high-energy photon in the signal signature is the photon candidate with highest transverse momentum, with minimum $\pT \geq \SI{30}{\GeV}$ and pseudorapidity in the range $|\eta|<1.37$ or $1.52 < |\eta| < 2.37$.
Photons are reconstructed from topological clusters of neighbouring cells in the electromagnetic calorimeter, not matched or matched to tracks in the inner detector depending on whether the photon underwent a conversion to $e^+e^-$ prior to reaching the calorimeter.
A tight photon identification selection, based on shower-shape criteria is applied to avoid selecting misidentified electrons or jets reconstructed as photons~\cite{EGAM-2018-01}.
Photon isolation requirements set an upper bound on the energy deposited in the calorimeter in a cone surrounding the photon, further helping to reject jets faking prompt photons.

Electron candidates must satisfy $p_{\mathrm{T}} > \SI{25}{GeV}$, $|\eta| < 2.47$, and tight identification criteria~\cite{EGAM-2018-01}.
Muon candidates must satisfy $p_{\mathrm{T}} > \SI{25}{GeV}$ and loose identification criteria~\cite{PERF-2015-10}.
For electrons, track-based isolation requirements depend on the electron's $p_{\mathrm{T}}$, using tracks within a variable cone size of up to \(\Delta R = 2\).
The calorimeter-based isolation requires that the sum of cluster transverse energies within a cone of \(\Delta R = 2\) is less than $\SI{3.5}{GeV}$.
For muons, the isolation requirement is that tracks within a cone of size $\Delta R = 0.2$ around the muon candidate track must sum to less than 1.25 GeV in \pT.
Any events containing isolated electrons or muons with $p_{\mathrm{T}} > \SI{25}{GeV}$ are vetoed.
Electron (muon) candidates are matched to the primary vertex by requiring that the significance of their transverse impact parameter $\dzero$ satisfies $|\dzero/\sigma(\dzero)|<5.0\ (3.0)$, where $\sigma(\dzero)$ is the measured uncertainty in $\dzero$, and by requiring that their longitudinal impact parameter $\zzsth$ satisfies $|\zzsth| < \SI{0.5}{mm}$.

To reduce instances of considering the same physics object as two different reconstructed objects, an overlap removal is performed after the kinematic criteria for object selection are met.
First, any jet within $\Delta R = 0.2$ of an electron is removed, and a jet with fewer than three associated tracks is removed if it is within $\Delta R = 0.2$ of a muon.
Any remaining electrons or muons found within $\Delta R = 0.4$ of a jet are then removed.
Then, any photon identified within $\Delta R = 0.4$ of an electron or muon is removed.
Finally, any jet that is within a region of $\Delta R = 0.4$ around a photon with $p_{\mathrm{T}} > \SI{30}{GeV}$ is removed, as it is likely to have been reconstructed from the same energy deposits as the photon candidate.

After the event selection criteria have been applied, candidate events are divided into two regions, which are used together in the classifier training and in the signal extraction fit.
The signal region (SR) contains most of the signal sensitivity, comprising events with $m_{bb}$ in the range 100~GeV~$\leq m_{bb} \leq$~150~GeV around the Higgs boson mass.
The control region (CR) includes events in the $m_{bb}$ sidebands outside this window, specifically $m_{bb}$ < 100 GeV and 150 GeV < $m_{bb} \leq$ 220 GeV.
It is primarily used to extract background normalizations and to constrain nuisance parameters representing correlated systematic uncertainties.

\begin{table}[htbp]
\caption{Trigger and offline event selection criteria and region definitions for the \(H(\rightarrow b\bar{b})jj + \gamma\) signature.
The \pt and \(|\eta|\) offline jet requirements are used to match trigger selections and \(b\)-tagging requirements.
Criteria in $m_{bb}$ define the control and signal regions.
}
\label{tab:eventselection}
\def\arraystretch{1.2}
\centering{
\begin{tabular}{ l | l | l }
\hline \hline
\multirow{4}{*}{Trigger} &     L1                    & \(\geq1\) photon with \(\pT > \SI{22}{\GeV}\) \\
\cline{2-3}
& \multirow{3}{*}{HLT}  & \(\geq1\) photon with \(\pT > \SI{25}{\GeV}\) \\
&         & \(\geq4\) jets with \(\pT>\SI{35}{\GeV}\) and \(|\eta|<4.9\) \\
&         &(or \(\geq3\) jets with \(|\eta|<4.9\) and \(\geq1\) $b$-tagged jet with \(|\eta|<2.5\))   \\
&                         & \(m_{jj}>\SI{700}{\GeV}\) \\
\hline
\multicolumn{2}{c|}{\multirow{5}{*}{Offline}} & \(\geq1\) photon with \(\pT>\SI{30}{\GeV}\) and \(|\eta|<1.37\) or \(1.52<|\eta|<2.37\) \\
\multicolumn{2}{c|}{}                         & \(\geq2\) $b$-tagged jets with \(\pT>\SI{40}{\GeV}\) and \(|\eta|<2.5\) \\
\multicolumn{2}{c|}{}                         & \(\geq2\) additional jets with \(\pT>\SI{40}{\GeV}\) and \(|\eta|<4.5\) \\
\multicolumn{2}{c|}{}                         & \(m_{jj} > \SI{800}{\GeV}\) \\
\multicolumn{2}{c|}{}                         & No electrons or muons with \(\pT>\SI{25}{\GeV}\)\\
\multicolumn{2}{c|}{}                         &  \(m_{bb}\leq\qty{220}{\GeV}\)\\
\hline
\multirow{2}{*}{Region selection} & Signal region & $\SI{100}{GeV} \leq m_{bb} \leq \SI{150}{GeV}$ \\
\cline{2-3}
& \multirow{1}{*}{Control region} & $m_{bb} < \SI{100}{GeV}$ and $\SI{150}{GeV} < m_{bb} \leq \SI{220}{GeV}$ \\
\hline \hline
\end{tabular}
}
\end{table}

\section{Multivariate analysis}
\label{sec:mva}

An improved multivariate analysis technique is key to this analysis.
An NN replaced the BDT used in the previous analysis because the NN showed better signal and background discrimination with identical inputs.
\textsc{Keras}~\cite{chollet2015keras}, a high-level interface of \textsc{TensorFlow}~\cite{tensorflow2015-whitepaper}, was used to train an NN to provide a classification score, distinguishing signal-like (high score) events from those resembling the non-resonant QCD $b\bar{b}\gamma jj$ or $c\bar{c}\gamma jj$ events constituting the main backgrounds (low score).
A two-fold method was used to train and evaluate the performance of the NN: two separate models with identical hyperparameters were trained on separate halves of the MC training samples, which contain both the control- and signal- region phase spaces.
Evaluating the MC events using the model that did not include those events during training allows the use of the full size of the MC samples for signal extraction.
Data and MC background samples that were not used in training are also split in half, and the two halves are evaluated by each of the two models and then recombined.

In order to optimize the choice of input variables and hyperparameters, a comparison metric was formed to assess the signal-background separation significance after successive training epochs.
The significance is estimated by binning the NN output score distribution into 20 bins, equal in width in score, and the weighted number of signal events is divided by the square root of the weighted number of background events ($S/\sqrt{B}$) in each bin.
The optimization metric is these $S/\sqrt{B}$ values summed in quadrature for all bins.
Many kinematic variables were considered as inputs; combinations of these variables were used to train individual models, which were evaluated using the defined metric to find the optimal set.
After optimization, the 14 input variables listed in Table \ref{tab:InputVariables} were selected, and no addition or substitution made to this grouping was found to produce better separation between signal and background.
In addition to the improvements gained from the use of an NN, this analysis profits from the use of the discriminating power of $m_{bb}$ and correlated variables, which could not be exploited in the previous analysis because it would sculpt the background parametrization.
The use of variables related to an event's jet multiplicity is also a new development, motivated by the signal's tendency, with mostly electroweak vertices, to have fewer jets than the QCD background.
The same optimization method was employed to select specific hyperparameter settings: each final NN model has 11 hidden layers with 256 neurons per layer in the bulk.
The ratio of the sums of weights of the background and signal training samples was also optimized, after which the signal acceptance contributed more than twice as much as the background rejection to the loss function.
To prevent overtraining, the training is stopped when the calculated loss function does not decrease significantly for five consecutive training epochs, which typically occurs within 25 epochs for the final selection of hyperparameters.

\begin{table}[h]
\caption{Kinematic variables used as inputs to the neural-network classifier.
This list is ordered in decreasing correlation between the input variable and the NN score, i.e. in decreasing order of importance in the NN training.}
\label{tab:InputVariables}
\centering
\begin{tabular}{p{0.15\linewidth} p{0.8\linewidth}}
\hline \hline
Input variable & Description \\
\hline
$n_\text{jets}$ & The total number of jets with $p_{\mathrm{T}} \geq \SI{40}{GeV}$ in the event \\
$p_{\mathrm{T}}^{\text{bal}}$~\cite{HIGG-2020-14} & Transverse momentum balance for selected final-state objects \\
$\Delta\phi(bb,jj)$ & Azimuthal angle between the momenta of the di-$b$-jet system and the two VBF-candidate jets \\
$m_{bb}$ & Invariant mass of the two leading $b$-tagged jets \\
$p_{\mathrm{T}}^{bb}$ & Transverse momentum of the two leading $b$-tagged jets \\
$m_{jj}$ & Invariant mass of the two VBF-candidate jets \\
$\cos\theta_C$ & Cosine of the angle between the two VBF-candidate jets and the two $b$-tagged jets planes in the centre-of-mass frame of the $bbjj$ system \\
$\min(\Delta R(b,\gamma))$ & The minimum angular distance between one of the $b$-tagged jets and the photon \\
$\Delta \eta (j, j)$ & Pseudorapidity difference between two VBF-candidate jets \\
$\Delta R(b_1,j_1)$ & Angular distance between the leading $b$-tagged jet and the leading VBF-candidate jet \\
$p_{\mathrm{T}}^{j_1}$ & Transverse momentum of the leading VBF-candidate jet \\
$\text{cen}(\gamma jj)$~\cite{HIGG-2020-14} & Centrality of the photon relative to the VBF-candidate jets \\
$\Delta \eta (b,b)$ & Pseudorapidity difference between the two leading $b$-tagged jets \\
$\eta_{j_5}$ & Pseudorapidity of a 5th jet (2 $b$-tagged jets, 2 VBF-candidate jets, plus this extra 5th jet); if no 5th jet is present, a value is assigned using a probability density matching the distribution of 5-jet events from the respective sample population \\
\hline\hline
\end{tabular}
\end{table}

\section{Signal and background composition and modelling}
\label{sec:modeling}

The previous analysis of this signature with the full Run-2 data relied on a fit to the invariant mass spectrum of the two $b$-tagged jets ($m_{bb}$), using an analytic shape for the background, with parameters determined from the fit to the data.
That approach introduced large systematic uncertainties from the potential bias arising from the choice of the model for the background.
This analysis adopts a binned likelihood fitting strategy to the NN score, based on signal and background templates derived from MC simulations, reducing modelling-related systematic uncertainties and benefitting from the use of a more discriminating variable as the fit observable.

Simulations and calculations of the SM cross-sections predict that non-resonant backgrounds are expected to contribute approximately 99\% of the total background.
Prior to any constraints from a fit to data, QCD multijet production with $b$-quarks in association with a photon ($b\bar{b}\gamma jj$) dominates with a 70\% contribution, multijet production with mistagged $c$-quarks in association with a photon ($c\bar{c}\gamma jj)$ contributes an estimated 18\%, and non-resonant production of top quarks ($t\bar{t}$) is estimated to contribute 11\% of the total expected background.
Non-resonant QCD production with light-flavour quarks in association with a photon ($l\bar{l}\gamma jj$) is conservatively estimated to contribute $<0.5\%$ of the total background yield based on the mistagging rate for light-flavour jets, so no template for this background source is included.

Although resonant backgrounds are estimated to contribute only $1\%$ of the total background, templates are still produced and included in the fit because the kinematics of resonant backgrounds are more similar to signal than to the non-resonant backgrounds.
The template for the background from $Z$ boson production in association with a photon and jets, $Z(\rightarrow b\bar{b})\gamma jj$, is built from the sum of two contributions, electroweak and QCD production, as described in Section~\ref{sec:mcsamples}.
The contribution from $W\gamma$ production is estimated to be only $1\%$ of that from $Z\gamma$ after selection, so no template is included in the fit for $W\gamma jj$.
Other Higgs production modes contribute negligibly to the selected phase space, due to the required associated photon and the VBF-specific event selections.
Nevertheless, templates are created for $ggH$ and $t\bar{t}H$ production, which is treated as background to the predominantly VBF Higgs signal.

While this analysis focuses on the VBF $H\gamma$ production mode of the $bb\gamma jj$ signal, $VH\gamma$ contributes an estimated 8.8\% of the signal sample at MC generator level, but it is expected to contribute less than 1\% after final event selection phase-space requirements.
Of the signal VBF events at generator level, about 92.3\% are estimated to be from Higgs boson production via $WW$-fusion, due to production via $ZZ$-fusion being suppressed by the interference effect mentioned in Section \ref{sec:intro}.

The templates for the $b\bar{b}\gamma jj$ and $c\bar{c}\gamma jj$ backgrounds use `truth-tagging' to benefit from as many of the MC events as possible, and tests were performed to ensure that the NN output distribution shapes are compatible with direct-tagged sample templates~\cite{ATL-PHYS-PUB-2022-041}.
These two backgrounds share a floating normalization factor, allowing their combined yield to be fitted directly to data.
The remaining background and signal templates use direct-tagged samples.
While truth-tagged MC signal events are used in the NN training to have ample training samples, the signal fit template is derived by evaluating the NN on only direct-tagged events.

As mentioned in Section \ref{sec:intro}, the fitting strategy extracts the signal directly from the NN output score distribution, and uses a control region and signal region defined by ranges in $m_{bb}$.
This fit relies on template distributions derived from MC, so it is important to ensure that the MC background simulation agrees well with data in the control region where there is little signal contamination.
In this control region, discrepancies between data and MC simulation are observed for several kinematic variables, particularly $n_\text{jets}$, $\min(\Delta R(b,\gamma))$, $\Delta \eta (j, j)$, and $m_{bb}$.
This was also observed in Ref.~\cite{HIGG-2020-14}, and is due to both the difficulty of modelling non-resonant multijet backgrounds and the low order at which the dominant MC samples were generated.

To achieve better agreement between data and MC events in the NN output distribution and to account for possible missing backgrounds such as events arising from jets improperly tagged as photons, reweighting scale factors for the two main background, $b\bar{b}\gamma jj$ and $c\bar{c} \gamma jj$, are extracted from distributions of the 14 NN input kinematic variables listed in Table \ref{tab:InputVariables}.
These backgrounds are reweighted together with identical kinematic scale-factor functions as follows.
After the MC-estimated $t\bar t$ contribution is subtracted, the data are compared with the predicted $b\bar{b}\gamma jj + c\bar{c}\gamma jj$ background, via the ratio of the normalized binned kinematic input-variable distributions in the $m_{bb}$-sideband control region.
To this ratio, a sixth-order polynomial is fitted successively and iteratively for each input variable, in the reverse order of their correlation with NN score.
The function is then applied to the MC $b\bar{b}\gamma jj$ and $c\bar{c} \gamma jj$ samples, in both the control region and signal regions, in order to scale these dominant backgrounds to better represent the data.
These kinematic reweighting scale factors account for any mismodelling of these non-resonant QCD backgrounds, as well as for any contamination from backgrounds not otherwise considered, such as jets misidentified as photons.

\begin{figure}
\centering
\subfigure[]{
\includegraphics[width=.48\textwidth]{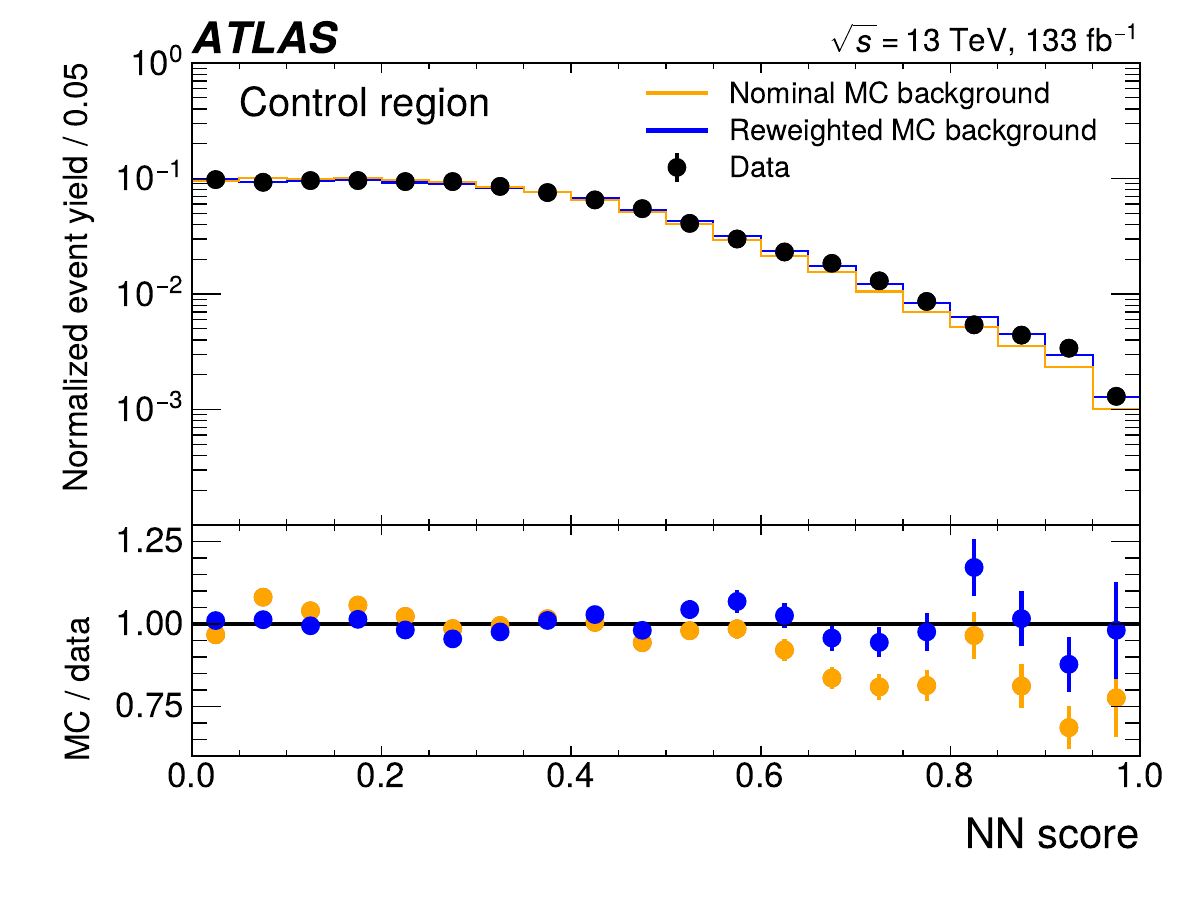}
}
\subfigure[]{
\includegraphics[width=.48\textwidth]{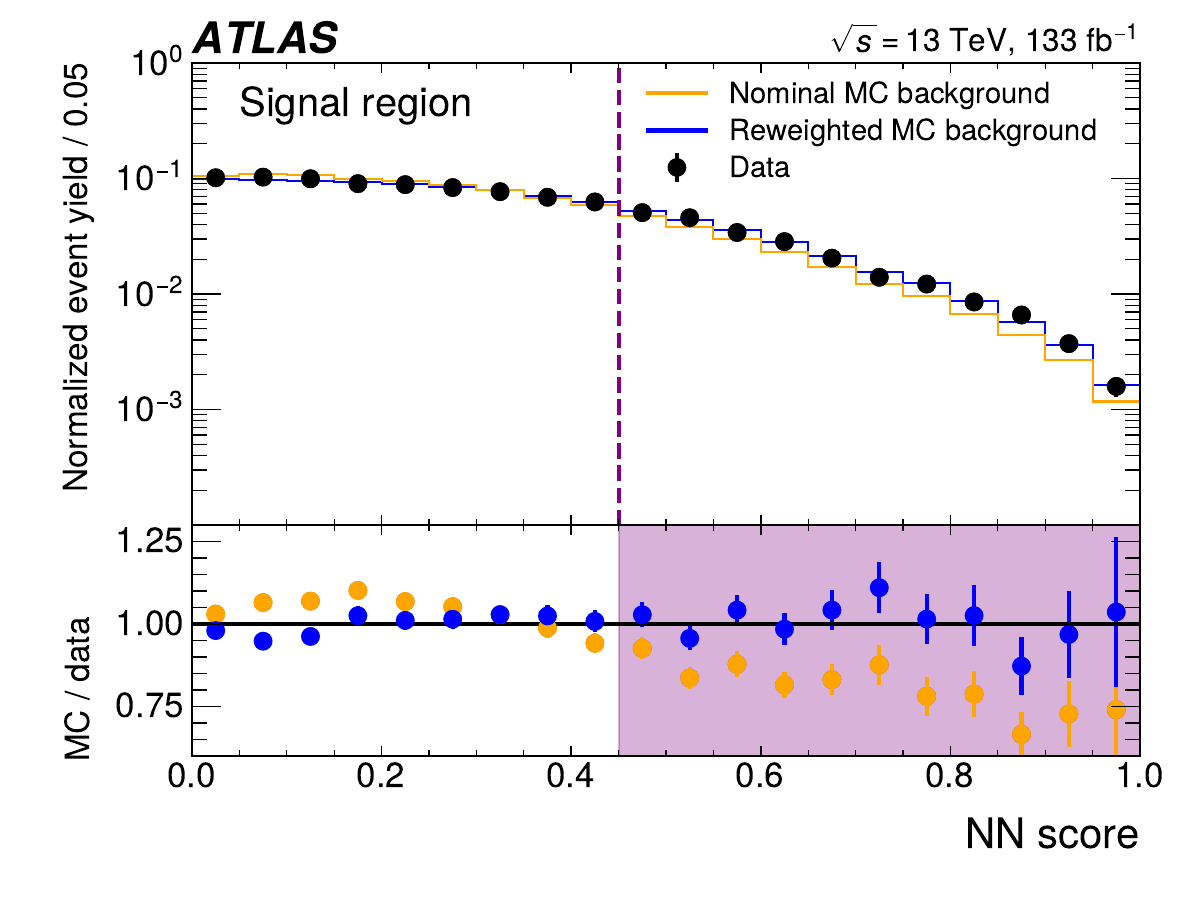}
}
\caption{NN output distributions for data compared with those for all MC backgrounds with and without kinematic reweighting.
Reweighting scale factors are extracted from comparisons between kinematic distributions for MC backgrounds and data in the (a) control region $m_{bb}$ sidebands.
The low score bins of the (b) signal region, which have negligible signal, are used for validation.
The purple dashed line marks the divide between the low-score SR bins, used for validation of the reweighting scale factors, and high-score bins containing most of the signal.
The total yields of MC and data events are individually normalized to 1, making this a direct comparison of NN output distribution shapes.}
\label{fig:mc_data_reweighting_nnoutput}
\end{figure}

The NN training is repeated after kinematic reweighting to obtain the final model, such that the modified event weights can affect the loss function and training results accordingly.
Despite the large number of compounding scale factors, the overall impact on the NN output is relatively small because their effects partially cancel.
Additionally, the reweighted result is largely immune to procedural changes, as reflected in the shape uncertainty due to this kinematic reweighting (see Section \ref{sec:systematics}).
Very good closure between MC and data distribution shapes after kinematic reweighting can be seen in the control region and the background-rich areas of the signal region, i.e. in the $m_{bb}$ SR window $100 \leq m_{bb} \leq 150$ GeV but with an NN score less than 0.6.
Figure \ref{fig:mc_data_reweighting_nnoutput} shows the NN output distributions for the MC backgrounds before and after kinematic reweighting in comparison with data in the control region and signal region.
In the CR and in the background-rich bins of the SR, the data and reweighted MC distributions agree within statistical uncertainties.

\section{Systematic uncertainties}
\label{sec:systematics}

Theory and experimental uncertainties are described with a total of 78 nuisance parameters that can modify the normalization and shape of the signal and background templates.
The effects of these nuisance parameters on the signal strength uncertainty ($\sigma(\mu_H)$) are categorized and summarized in Table~\ref{tab:Breakdown_unblinded}.
The grouped effect of each uncertainty category is calculated by performing a fit with the category's corresponding nuisance parameters fixed to their best-fit values, and taking the difference in quadrature between $\sigma(\mu_H)$ from the nominal fit and from this fit as the impact of that group of systematic uncertainties.
Template smoothing is employed in these final-fit variations to reduce the inflation of uncertainties due to correlated statistical fluctuations.

\begin{table}[h]
\caption{
The observed impacts of the statistical, theory, and experimental uncertainties on the signal strength uncertainty, as extracted from the final fit to observed data.
Data statistical uncertainty is the single largest contribution to the total uncertainty.}
\label{tab:Breakdown_unblinded}
\begin{center}
\begin{tabular}{lcc}
\hline \hline
Source of uncertainty &  \(+\sigma(\mu_H)\)& \(-\sigma(\mu_H)\)\\
\hline
Statistical\\
\quad Data sample size & \(+0.57\) & \(-0.56\) \\
\quad MC background templates & \(+0.25\) & \(-0.22\)  \\

\hline
Theory and modelling\\
\quad Kinematic reweighting & \(+0.07\) & \(-0.06\) \\
\quad MC modelling & \(+0.19\) & \(-0.15\)  \\
\quad Background normalization & \(+0.13\) & \(-0.08\) \\

\hline
Experimental\\
\quad Luminosity & \(+0.004\) & \(-0.006\) \\
\quad Jet + photon selection & \(+0.15\) & \(-0.11\)  \\
\quad \(b\)-tagging &  \(+0.03\) & \(-0.03\)   \\

\hline \hline
Total & \(+0.69\)& \(-0.64\) \\
\quad Total MC statistical & \(+0.25\)& \(-0.22\) \\
\quad Total theory & \(+0.24\)& \(-0.18\) \\
\quad Total experimental & \(+0.16\)& \(-0.12\) \\
\hline \hline
\end{tabular}
\end{center}
\end{table}

The dominant source of signal strength uncertainty is the statistical uncertainty, which contributes \(\sigma(\mu_H) = 0.57\).
This is due to the small \(H+\gamma\) production cross section at \(\sqrt{s}=\SI{13}{TeV}\).
Nevertheless, much attention was paid to reducing the experimental and modelling uncertainties as much as possible relative to the previous analysis.
To reduce the statistical uncertainties of the background templates, new and enlarged \(b\bar{b}\gamma jj\) and \(c\bar{c}\gamma jj\) samples were simulated; these reduced the MC statistical uncertainty contributions to a total of \(\sigma(\mu_H) = 0.25\).

Theory uncertainties arising from the lack of higher-order corrections, the PDF and $\alpha_s$, and the parton shower generator are considered for the signal and major backgrounds.
To estimate the contributions from missing higher-order corrections in the calculations, an envelope of the effects of varying the chosen renormalization and factorization scale values is used.
The dominant normalization uncertainties of the $b\bar{b}\gamma jj$ and $c\bar{c}\gamma jj$ backgrounds come from this variation of the renormalization scale $\mu_R$ and the factorization scale $\mu_F$.
MC events were reweighted in accord with changes in these QCD scales, leading to a change of roughly $+60\%$ and $-40\%$ in the yield for each sample, independent of NN score.
The effects of a seven-point variation, $(\{\mu_R,\mu_F\}\times\{0.5,0.5\},\{1.0,0.5\},\{0.5,1.0\},\{1.0,1.0\},\{1.0,2.0\},\{2.0,1.0\},\{2.0,2.0\})$, is taken as the systematic uncertainty, including both the cross-section and acceptance uncertainties.
Shape effects from these scale variations are not large, but are included and treated as being uncorrelated with the normalization effects to avoid a degeneracy between these nuisance parameters and the floating normalization factor.
The scale-uncertainty nuisance parameters for $b\bar{b}\gamma jj$ and $c\bar{c}\gamma jj$ are assumed to be uncorrelated as well, despite their production processes being similar, and in doing so, the ratio of the yields of the two samples is given a large range of freedom.

The uncertainties due to missing higher orders for the signal are evaluated in the same way as for the background, but they are found to be much smaller, about $1\%$, and correlation between normalization and shape components is allowed in the fit.
For the $t\bar{t}$ background, the uncertainty in the cross-section estimated at NNLO in QCD~\cite{Martin:2009bu} is used to account for QCD scale-variation effects, and thus an overall relative uncertainty of $^{+2.5\%}_{-3.6\%}$ is assigned to the $t\bar{t}$ background.
For the small $Z\gamma jj$ background, a conservative estimate of $\pm16\%$ is adopted for the overall magnitude of scale-variation effects, matching the uncertainty estimated in Ref.~\cite{EXOT-2021-17}.

Systematic uncertainties due to the choice of PDF and $\alpha_s$ are estimated by reweighting probability densities at the event level.
For signal, 42 members of the set PDF4LHC15 are considered, including 40 PDF eigenvariations along with $\alpha_s$ variations upward and downward by 1 standard deviation.
The uncertainty is calculated as the sum in quadrature of the effects of these variations~\cite{Butterworth:2015oua}, which leads to an uncertainty of $2.2\%$ for signal.
For $b\bar{b}\gamma jj$ and $c\bar{c}\gamma jj$, 100 PDF variations of the set PDF4LHC21 are considered, and the uncertainty is computed as the standard deviation of the yields from each variation.
This PDF uncertainty is combined in quadrature with the change in yield from the average of the upward and downward $\alpha_s$ variations.
The resulting PDF and $\alpha_s$ choice uncertainty is $3.2\%$ for $b\bar{b}\gamma jj$ and $3.3\%$ for $c\bar{c}\gamma jj$.
The equivalent uncertainty for $t\bar{t}$ production is $2.5\%$, again taken from the uncertainty in the best prediction of the NNLO cross-section~\cite{botje2011pdf4lhcworkinggroupinterim}.
PDF and $\alpha_s$ uncertainties are not considered directly for the other backgrounds, as such small variations on the rarer backgrounds have negligible impact on the signal measurement.
However, for the minor Higgs boson production modes that are treated as backgrounds in this analysis, an overall normalization uncertainty of $20\%$ is introduced to conservatively cover any modelling uncertainties.

The largest shape variation from a single systematic uncertainty source comes from the choice of parton shower generator, giving bin-by-bin variations as large as 5\%.
For the dominant $b\bar{b}\gamma jj$ and $c\bar{c}\gamma jj$ backgrounds, alternative samples were produced using \MGNLO and \HERWIG, instead of the \MGNLO and \PYTHIA combination used in obtaining the nominal templates.
These samples were produced at MC truth-level, so a method was designed to extract the truth-level effects of differences between \PYTHIA and \HERWIG showering and transfer those effects to modifications of the fully simulated NN output distributions.
Scale factors are extracted from the ratios of truth-level \PYTHIA and \HERWIG kinematic distributions for each of the 14 NN input variables (similar to the process for kinematic reweighting discussed in Section \ref{sec:modeling}).
These scale factors are then applied to the fully simulated and kinematically reweighted \PYTHIA NN outputs to obtain a facsimile of a fully simulated \HERWIG NN outputs.
The effects of these representative parton shower variations are included as systematic uncertainties with two-sided symmetrization.
Separate nuisance parameters were introduced for the signal region and control region for these dominant backgrounds, because the NN output distribution shape is affected differently in the two regions.

For signal, the alternative shower samples were generated with \HERWIG[7] like the nominal samples but with variations of the hard scale factor by factors of two, which affects the maximum allowed transverse momentum for shower emissions~\cite{Bellm:2015jjp}.
The method used for transfer from truth-level to proxy fully simulated distributions is the same as described for the two main non-resonant backgrounds, and it results in bin-by-bin variations as large as 20\% in signal-sensitive regions.
For $t\bar{t}$ production, hard-scatter (matrix element) and parton shower uncertainties are estimated by comparing the nominal samples directly with fully simulated samples generated with \MGNLO[5] replacing \POWHEG and \HERWIG[7] replacing \PYTHIA[8].
The parton shower variation changes the $t\bar{t}$ sample normalization by almost $10\%$.
The effects of the matrix element variation are small in the control region, changing the $t\bar{t}$ normalization by $0.6\%$, but larger in the signal region, with a normalization change of $5.3\%$.

The kinematic reweighting of the dominant $b\bar{b}\gamma jj$ and $c\bar{c}\gamma jj$ backgrounds also has associated systematic uncertainties.
The reweighting is performed by extracting scale factors from the ratio of the sum of these two simulated backgrounds to the $t\bar{t}$-subtracted data.
Thus, the systematic uncertainty is derived by considering how the result of this reweighting changes if the $t\bar{t}$ contribution is overestimated or underestimated within its theoretical uncertainties.
Since the scale factors are extracted concurrently for $b\bar{b}\gamma jj$ and $c\bar{c}\gamma jj$ directly in the control region and extrapolated to the signal region, uncertainty is considered only in the signal region and is correlated between the two backgrounds.
Due to the robustness of the kinematic reweighting procedure, these $t\bar{t}$ contribution variations have little impact on the NN output distribution's shape.
In the unique case of $m_{bb}$, for which the phase space covered in the signal region is not represented in the control region, the entire reweighting is taken as an uncertainty: a systematic variation without the scale factors derived for $m_{bb}$ is included in the signal region, and the associated nuisance parameter ranks among the highest in terms of impact to the signal strength uncertainty.
The combination of these two kinematic reweighting uncertainty nuisance parameters gives a $6-7\%$ impact on the signal strength uncertainty.

Experimental systematic uncertainties arise from uncertainties in the integrated luminosity, jet energies and resolutions, electromagnetic object identification and isolation, and $b$-jet tagging efficiency.
These systematic uncertainties are correlated between samples and fit regions via shared nuisance parameters.
Due to the uncertainty in the measurement of the integrated luminosity collected by ATLAS~\cite{DAPR-2021-01}, an overall normalization uncertainty of 0.83\% is applied to every template sample, but this has negligible impact on the final signal strength uncertainty.
Systematic uncertainties arising from jet energy scale (JES) and resolution (JER) uncertainties~\cite{JETM-2018-05} dominate the experimental category with a \(13\%\) contribution to the signal strength uncertainty.
Experimental uncertainties arising from the identification and isolation of photons and electrons, as well as their energy and its resolution~\cite{EGAM-2018-01}, are also considered but have negligible impact.
Uncertainties in $b$-jet tagging efficiency include effects for both online and offline selections~\cite{FTAG-2018-01, FTAG-2020-08, FTAG-2019-02}.
The overall background uncertainty arising from $b$-tagging requirements is about $2-3\%$ and fairly independent of NN score for all samples, but variations in both shape and normalization are considered for the signal and dominant background samples.
In all, $b$-jet tagging uncertainties contribute \(3\%\) to the signal strength uncertainty.

\section{Results}
\label{sec:result}

To extract the signal strength, a binned maximum-likelihood fit is performed directly on the neural-network classifier's output distribution using TRExFitter, which is built from \textsc{HistFactory}~\cite{Cranmer:2012sba} and \textsc{RooFit}~\cite{RooFit}.
The signal strength and the floating normalization factor for the sum of the two main non-resonant backgrounds, \(b\bar{b}\gamma jj\) and \(c\bar{c}\gamma jj\), are extracted~\cite{Rolke:2004mj} by a simultaneous fit of the control and signal regions, which are defined by the invariant mass of the two leading \(b\)-tagged jets as described in Section \ref{sec:modeling}.

\begin{figure}
\begin{centering}
\subfigure[]{
\includegraphics[width=0.45\textwidth]{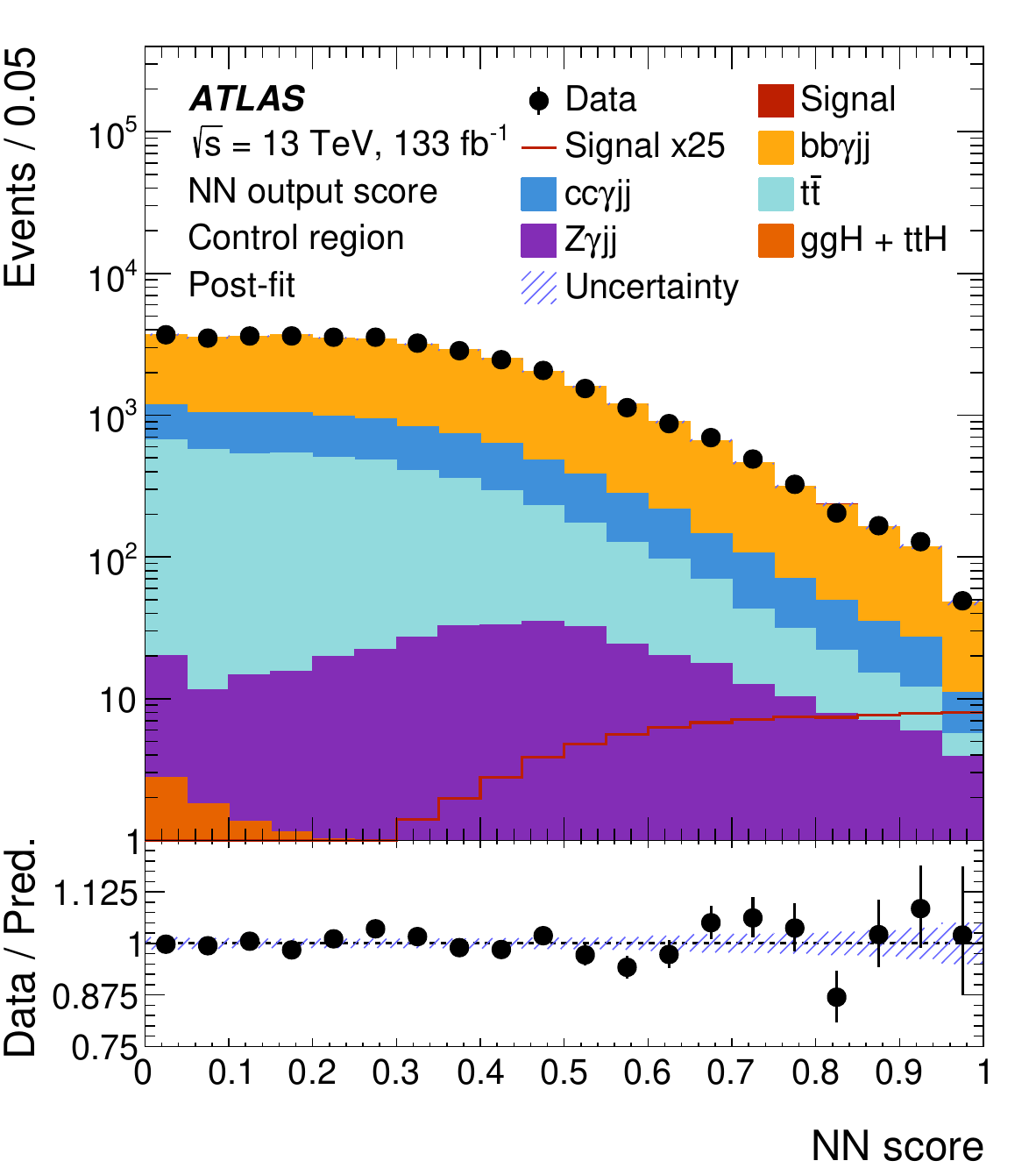}
}
\subfigure[]{
\includegraphics[width=0.45\textwidth]{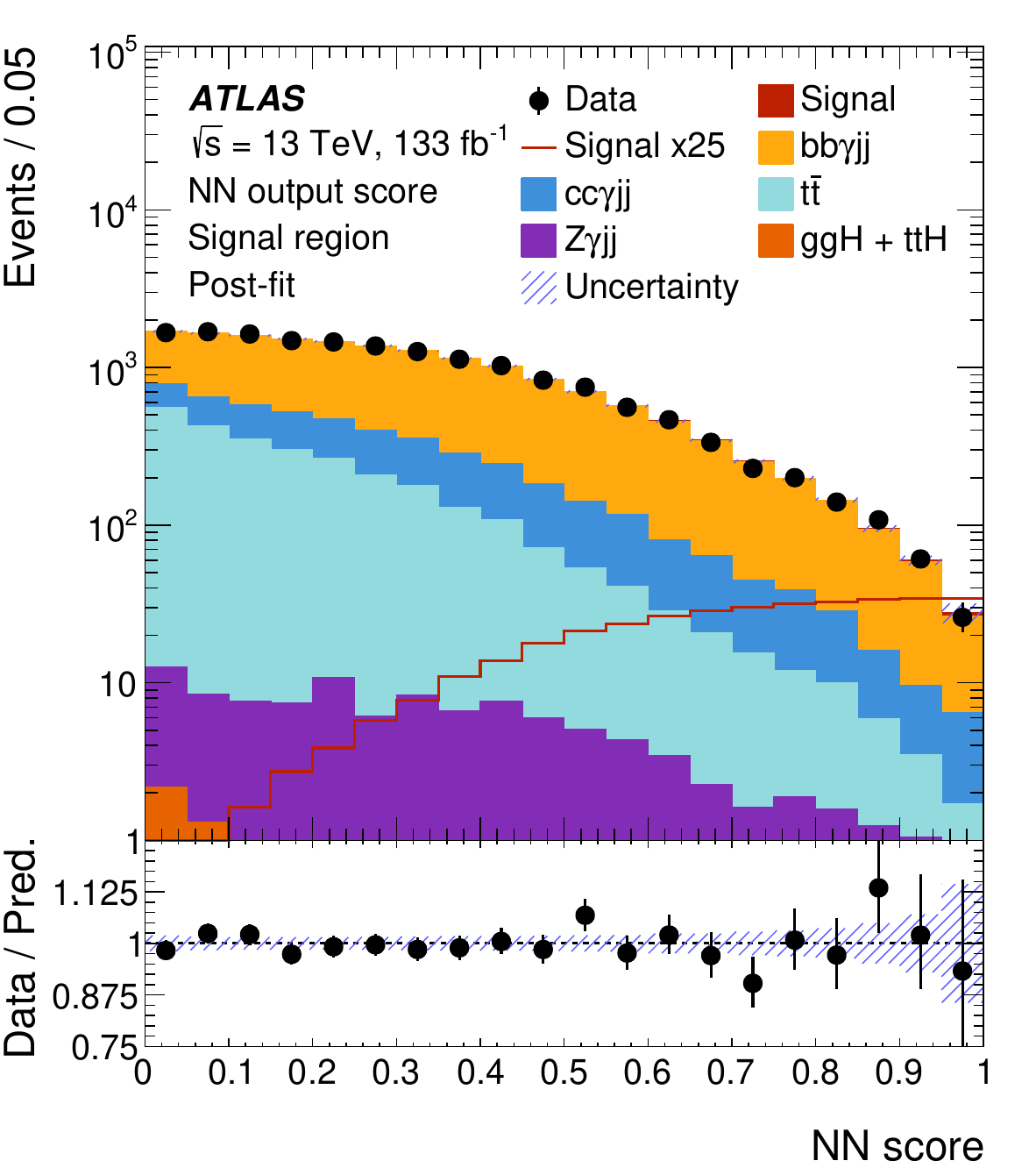}
}
\caption{Post-fit plots for the observed dataset, showing (a) the control region and (b) the signal region.
Background and signal templates are shown with all normalization factors and post-fit nuisance parameter pulls incorporated.
Data bin yields are represented with black markers, with error bars representing statistical uncertainties.
The red line represents the post-fit signal scaled by a factor of 25, while the stacked red signal histogram, which is only visible at high NN score in the signal region, is normalized to the fitted signal strength of \(\mu=0.2\).
The ratio plots in the lower panels display the bin-by-bin ratio of the data to the post-fit total of signal and background yields.
Post-fit uncertainties are depicted with a blue hashed band in both the yield and ratio plots.
The NN score is renormalized in the post-fit distributions to give regular bin widths, so the binning does not reflect the raw score optimized binning, which is finer in high score signal region bins and low score control region bins.}
\label{fig:postfit_plots}
\end{centering}
\end{figure}

Both the control region and signal region have 20 NN score bins with widths optimized independently for each region to maximize the signal sensitivity while keeping the total statistical uncertainties low.
The optimization results in very fine binning in the high score range of the signal region, where the signal purity is highest, as well as fine binning in the low score range of the control region, where the number of background events is large.
The highest-score bins of the signal region have by far the highest signal purity and sensitivity in the fit.
The post-fit NN score distributions in the control region and signal region are shown in Figure~\ref{fig:postfit_plots}.
The pre-fit and post-fit yields for the MC-simulated backgrounds and signal are listed in Table~\ref{tab:Yields} together with the number of data events.

The Higgs boson signal strength is defined relative to the Standard Model prediction.
An inclusive signal strength of \(\mu_H = 0.2 \pm 0.7\) is observed, compared to an expected value of \(1.0 \pm 0.7\).
This corresponds to an observed significance of \(0.3\) standard deviations, compared to an expected significance of \(1.5\) standard deviations.
The 95\% confidence level exclusion limit for the VBF H$\gamma$jj signal cross-section is observed to be \(86.5\,\)fb.
A likelihood ratio comparison between the observed fit and a saturated model, which has complete freedom for a perfect fit to data, gives a goodness-of-fit \(p\)-value of \(30\%\).
The final fit's normalization of the dominant backgrounds is about 85\% of the leading-order MC estimate.
The measured signal strength is compatible with the previous result, with a \(p\)-value of \(37\%\).
This probability is calculated without considering correlations because the overlap of data events in the signal-sensitive regions of the two analyses is less than \(5\%\).

\begin{table}
\caption{Pre-fit and post-fit signal and background yields and data yields in the control region, and in the signal region in the sum of the 17 lowest-NN-score bins and individually in the three highest-NN-score bins.
The total background yield and its uncertainty are given post-fit.}
\label{tab:Yields}
\begin{center}
\begin{tabular}{|c|r|r|r|r|r|r|r|r|r|r|}
\hline
Yields & \multicolumn{2}{|c|}{CR}      & \multicolumn{2}{c|}{SR Bins 0-16} & \multicolumn{2}{c|}{SR Bin 17} & \multicolumn{2}{c|}{SR Bin 18} & \multicolumn{2}{c|}{SR Bin 19} \\
& \multicolumn{1}{|c|}{Pre} & \multicolumn{1}{|c|}{Post} & \multicolumn{1}{|c|}{Pre} & \multicolumn{1}{|c|}{Post} & \multicolumn{1}{|c|}{Pre} & \multicolumn{1}{|c|}{Post} & \multicolumn{1}{|c|}{Pre} & \multicolumn{1}{|c|}{Post} & \multicolumn{1}{|c|}{Pre} & \multicolumn{1}{|c|}{Post} \\ \hline
\multicolumn{1}{|c|}{$b\bar{b}\gamma jj$}       & 30600 & 27400 & 12400 & 11100   & 86 & 78      & 55 & 49   & 23 & 20     \\ \hline
\multicolumn{1}{|c|}{$c\bar{c}\gamma jj$}       & 7400 & 5100   & 3200 & 2200     & 15 & 10      & 9 & 6     & 7 & 5       \\ \hline
\multicolumn{1}{|c|}{$t\bar{t}$}                & 4600 & 4800   & 2600 & 2700     & 4 & 5        & 2 & 2     & 1 & 1       \\ \hline
\multicolumn{1}{|c|}{$Z\gamma jj$}              & 400 & 400     & 100 & 100       & 1 & 1        & 1 & 1     & 0.4 & 0.4       \\ \hline
\multicolumn{1}{|c|}{$ggH + t\bar{t}H$}         & 10 & 10       & 8 & 8           & <0.1 & <0.1  & <0.1 & <0.1  & <0.1 & <0.1  \\ \hline \hline
\multicolumn{1}{|c|}{Total background}          & \multicolumn{2}{c|}{$37700 \pm 130$} & \multicolumn{2}{c|}{$16100 \pm 76$} & \multicolumn{2}{c|}{$94 \pm 5$} & \multicolumn{2}{c|}{$59 \pm 4$} & \multicolumn{2}{c|}{$27 \pm 4$} \\ \hline
\multicolumn{1}{|c|}{Signal}                    & 14 & 3    & 41 & 9         & 6 & 1      & 6 & 1      & 6 & 1      \\ \hline
\multicolumn{1}{|c|}{Data}                      & \multicolumn{2}{c|}{37701}   & \multicolumn{2}{c|}{16070}        & \multicolumn{2}{c|}{108}       & \multicolumn{2}{c|}{61}        & \multicolumn{2}{c|}{26}        \\ \hline
\end{tabular}
\end{center}
\end{table}

\begin{figure}
\centering
\subfigure[]{
\includegraphics[width=0.4\linewidth]{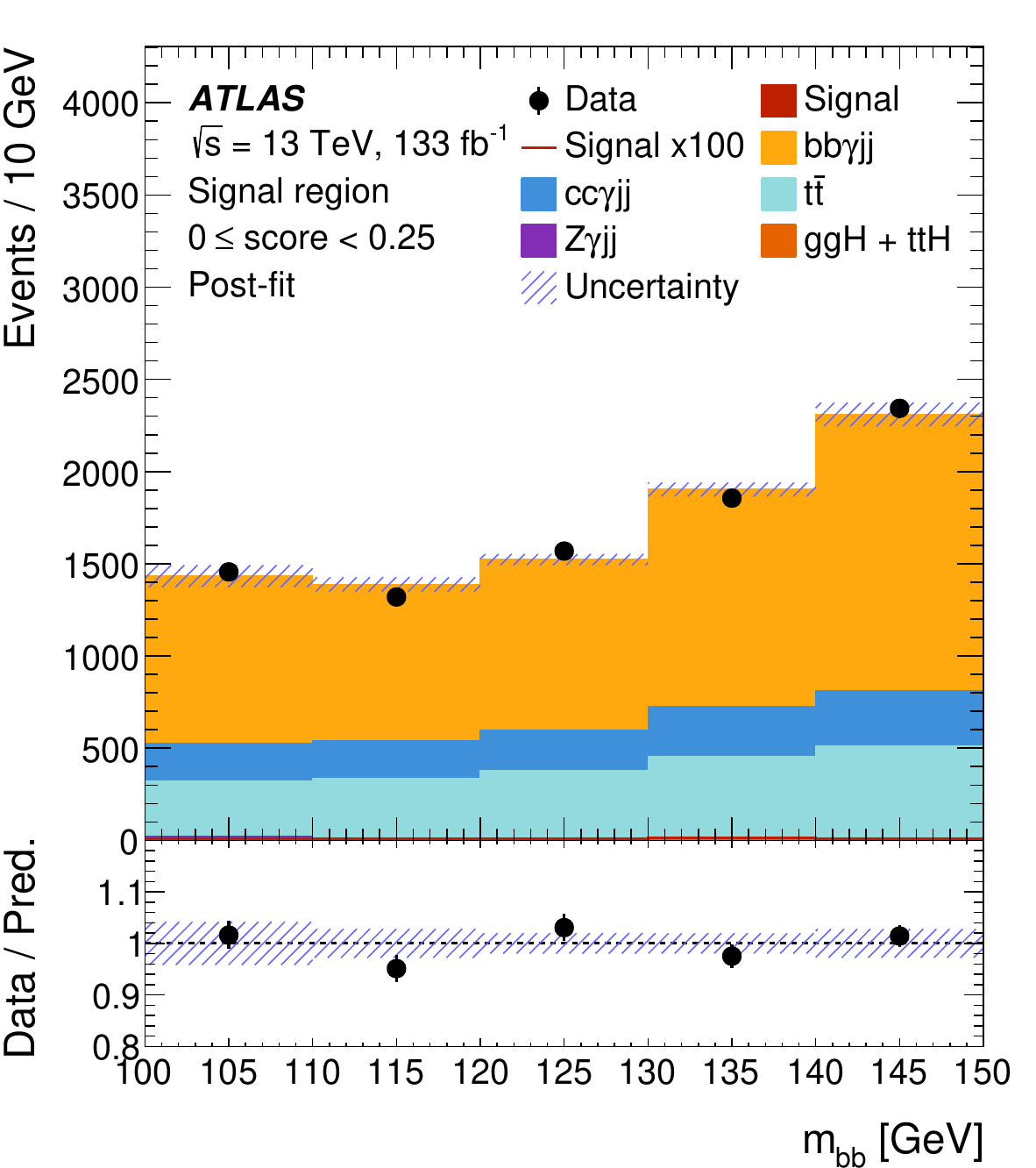}
}
\subfigure[]{
\includegraphics[width=0.4\linewidth]{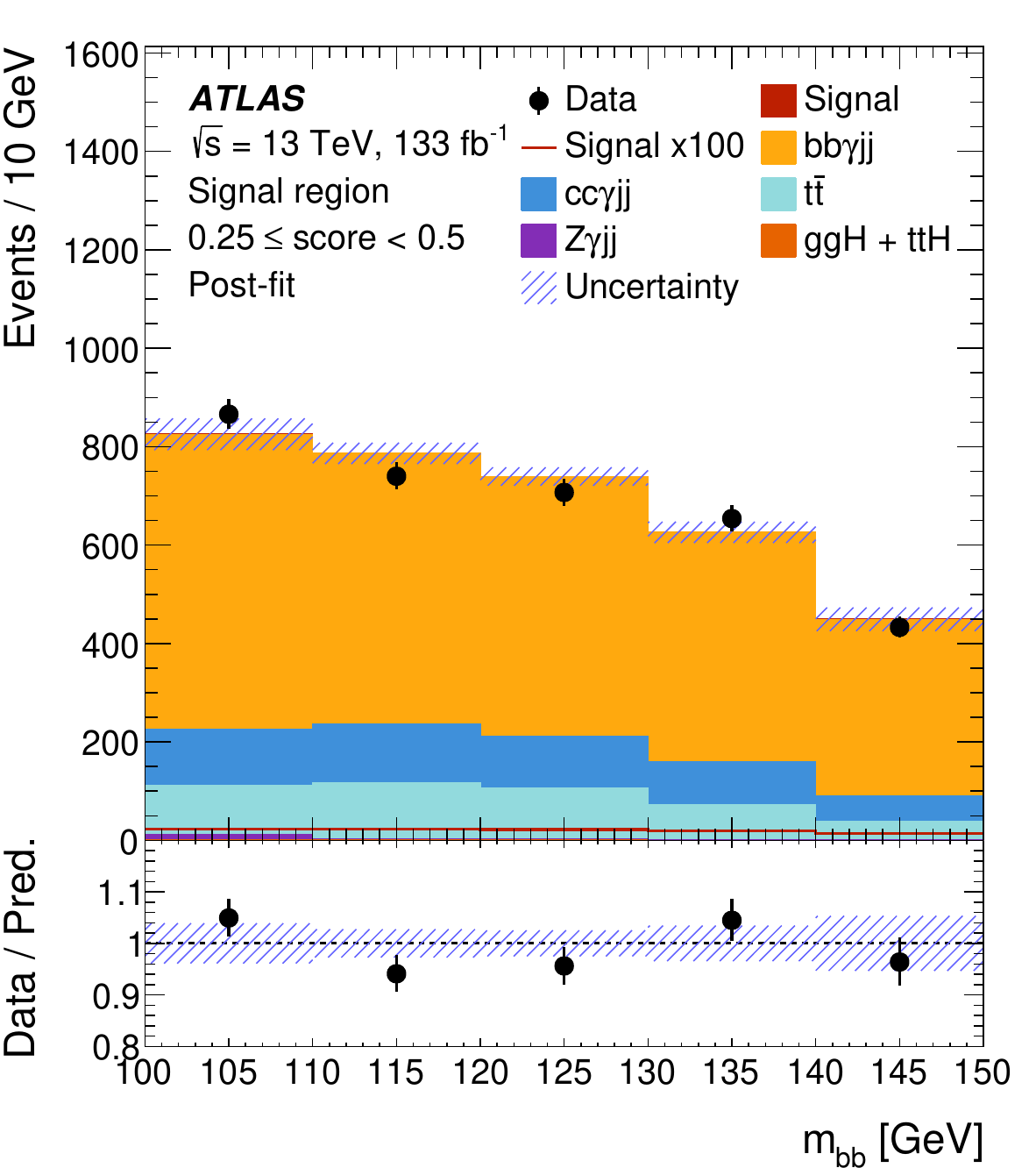}
}
\subfigure[]{
\includegraphics[width=0.4\linewidth]{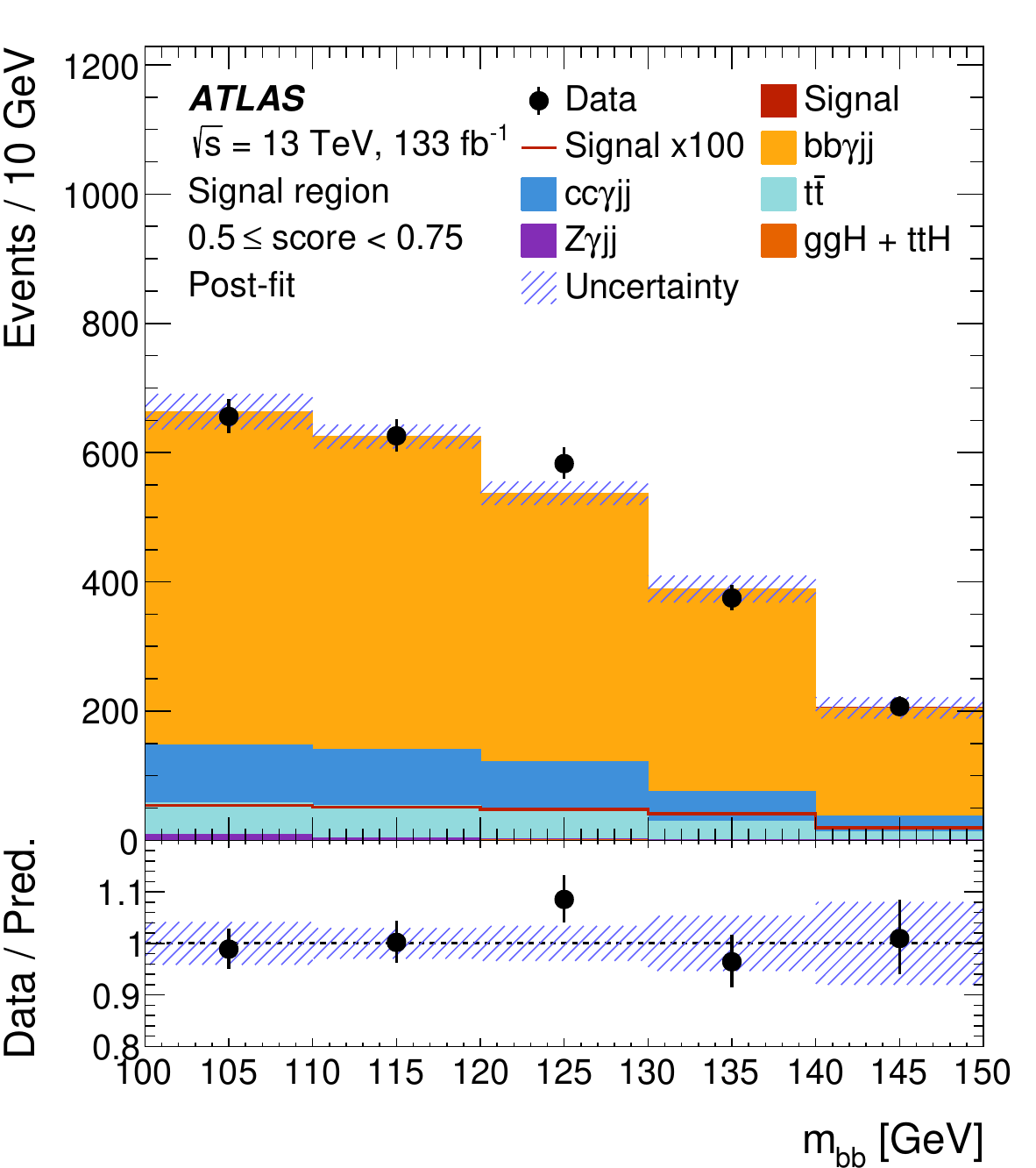}
}
\subfigure[]{
\includegraphics[width=0.4\linewidth]{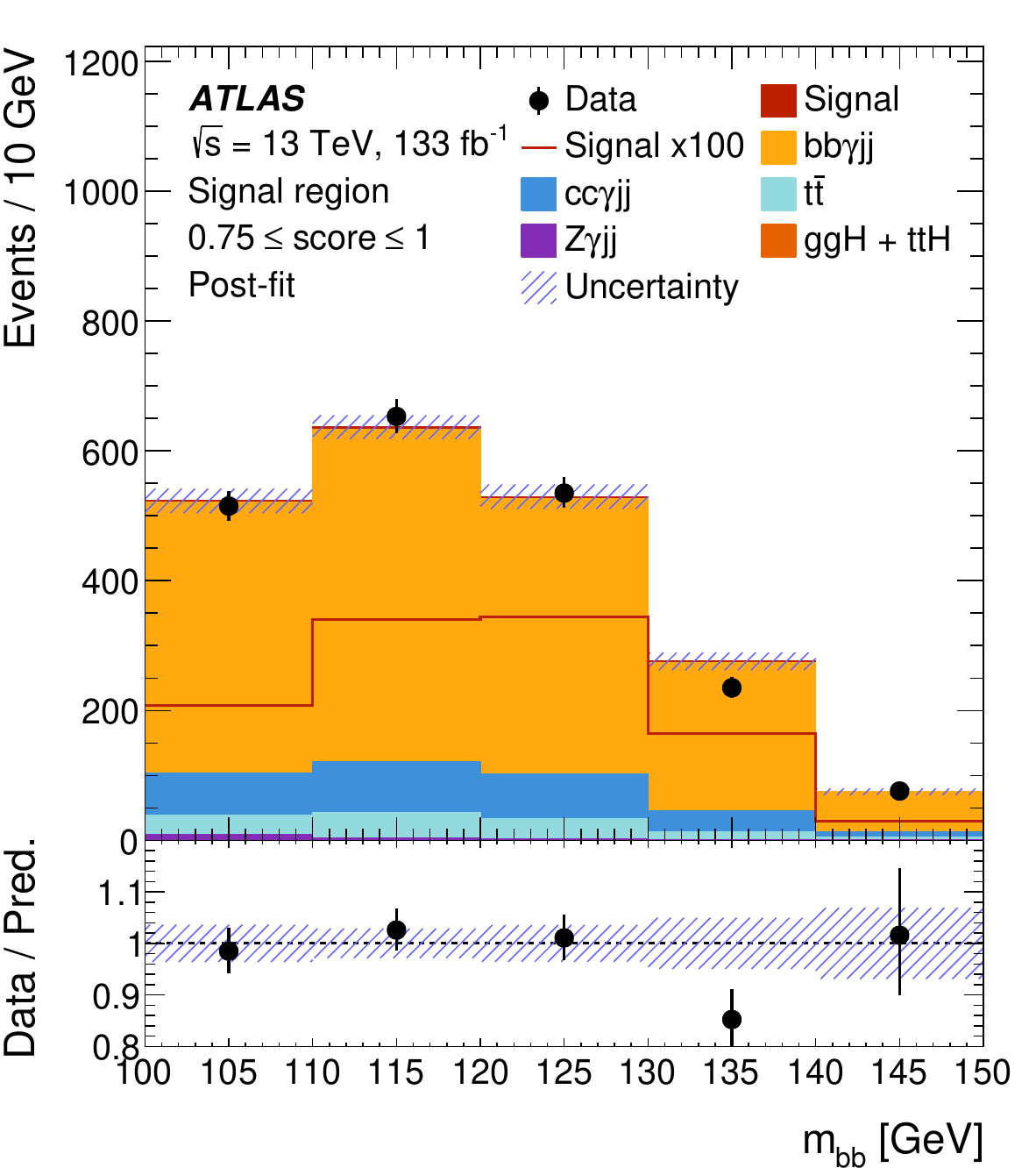}
}
\caption{Post-fit signal region \(m_{bb}\) distributions in slices of NN score.
Stacked histograms represent post-fit MC background and signal, and the black markers with error bars represent data.
The red lines represent post-fit signal distributions scaled by a factor of 100.
The stacked red histogram, which can only be visualized in some bins of the highest NN score slice, is normalized to the post-fit signal strength, \(\mu=0.2\).
The lower panels show the ratio of data to post-fit MC signal plus background.
Post-fit uncertainties are depicted by blue hashed lines.}
\label{fig:mBB_sliced}
\end{figure}

To further confirm that the background modelling using the MC-derived templates is acceptable, kinematic distributions are compared with data in the signal region, in slices of NN score, with the propagation of post-fit pulls from the fit to NN score.
Figure~\ref{fig:mBB_sliced} shows the di-\(b\)-jet invariant mass spectrum in four NN score slices of the signal region.
The post-fit MC and data spectra agree within their uncertainties, and no divergence in agreement appears at high score.

\FloatBarrier

\section{Conclusion}

A search was conducted for associated production of a Higgs boson and a high-energy photon.
The full Run-2 dataset from 13 TeV proton--proton collisions collected by the ATLAS detector at the LHC is examined, corresponding to an integrated luminosity of \(\SI{133}{\ifb}\).
The search focuses on the \(H(\rightarrow b\bar{b})+\gamma\) final state, the dominant Higgs boson decay mode, and the photon requirement greatly reduces the multijet background.
The photon requirement also makes vector-boson fusion the dominant Higgs boson production mode, with events characterized by the additional presence of two well-separated VBF-candidate jets having a high invariant mass.
This analysis improves on a previous one in this channel by using updated and extended Monte Carlo background samples, finely optimized multivariate analysis techniques for signal--background separation, and a novel data-fitting approach to measure the signal strength.
With these improvements, a signal significance of \(1.5 \sigma\) is expected, nearly 50\% higher than the expected significance in the previous full Run-2 analysis.
A Higgs boson signal strength of \(0.2 \pm 0.7\) is measured, with an observed significance of \(0.3\sigma\).

\section*{Acknowledgements}
%

%
%

%
%

We thank CERN for the very successful operation of the LHC and its injectors, as well as the support staff at
CERN and at our institutions worldwide without whom ATLAS could not be operated efficiently.

The crucial computing support from all WLCG partners is acknowledged gratefully, in particular from CERN, the ATLAS Tier-1 facilities at TRIUMF/SFU (Canada), NDGF (Denmark, Norway, Sweden), CC-IN2P3 (France), KIT/GridKA (Germany), INFN-CNAF (Italy), NL-T1 (Netherlands), PIC (Spain), RAL (UK) and BNL (USA), the Tier-2 facilities worldwide and large non-WLCG resource providers. Major contributors of computing resources are listed in Ref.~\cite{ATL-SOFT-PUB-2025-001}.

We gratefully acknowledge the support of ANPCyT, Argentina; YerPhI, Armenia; ARC, Australia; BMWFW and FWF, Austria; ANAS, Azerbaijan; CNPq and FAPESP, Brazil; NSERC, NRC and CFI, Canada; CERN; ANID, Chile; CAS, MOST and NSFC, China; Minciencias, Colombia; MEYS CR, Czech Republic; DNRF and DNSRC, Denmark; IN2P3-CNRS and CEA-DRF/IRFU, France; SRNSFG, Georgia; BMFTR, HGF and MPG, Germany; GSRI, Greece; RGC and Hong Kong SAR, China; ICHEP and Academy of Sciences and Humanities, Israel; INFN, Italy; MEXT and JSPS, Japan; CNRST, Morocco; NWO, Netherlands; RCN, Norway; MNiSW, Poland; FCT, Portugal; MNE/IFA, Romania; MSTDI, Serbia; MSSR, Slovakia; ARIS and MVZI, Slovenia; DSI/NRF, South Africa; MICIU/AEI, Spain; SRC and Wallenberg Foundation, Sweden; SERI, SNSF and Cantons of Bern and Geneva, Switzerland; NSTC, Taipei; TENMAK, T\"urkiye; STFC/UKRI, United Kingdom; DOE and NSF, United States of America.

Individual groups and members have received support from BCKDF, CANARIE, CRC and DRAC, Canada; CERN-CZ, FORTE and PRIMUS, Czech Republic; COST, ERC, ERDF, Horizon 2020, ICSC-NextGenerationEU and Marie Sk{\l}odowska-Curie Actions, European Union; Investissements d'Avenir Labex, Investissements d'Avenir Idex and ANR, France; DFG and AvH Foundation, Germany; Herakleitos, Thales and Aristeia programmes co-financed by EU-ESF and the Greek NSRF, Greece; BSF-NSF and MINERVA, Israel; NCN and NAWA, Poland; La Caixa Banking Foundation, CERCA Programme Generalitat de Catalunya and PROMETEO and GenT Programmes Generalitat Valenciana, Spain; G\"{o}ran Gustafssons Stiftelse, Sweden; The Royal Society and Leverhulme Trust, United Kingdom.

In addition, individual members wish to acknowledge support from Armenia: Yerevan Physics Institute (FAPERJ); CERN: European Organization for Nuclear Research (CERN DOCT); Chile: Agencia Nacional de Investigaci\'on y Desarrollo (FONDECYT 1230812, FONDECYT 1240864, Fondecyt 3240661, Fondecyt Regular 1240721); China: Chinese Ministry of Science and Technology (MOST-2023YFA1605700, MOST-2023YFA1609300), National Natural Science Foundation of China (NSFC - 12175119, NSFC 12275265, NSFC-12075060); Czech Republic: Czech Science Foundation (GACR - 24-11373S), Ministry of Education Youth and Sports (ERC-CZ-LL2327, FORTE CZ.02.01.01/00/22\_008/0004632), PRIMUS Research Programme (PRIMUS/21/SCI/017); EU: H2020 European Research Council (ERC - 101002463); European Union: European Research Council (BARD No. 101116429, ERC - 948254, ERC 101089007), Horizon 2020 Framework Programme (MUCCA - CHIST-ERA-19-XAI-00), European Union, Future Artificial Intelligence Research (FAIR-NextGenerationEU PE00000013), Italian Center for High Performance Computing, Big Data and Quantum Computing (ICSC, NextGenerationEU); France: Agence Nationale de la Recherche (ANR-20-CE31-0013, ANR-21-CE31-0013, ANR-21-CE31-0022, ANR-22-EDIR-0002, ANR-24-CE31-0504-01); Germany: Baden-Württemberg Stiftung (BW Stiftung-Postdoc Eliteprogramme), Deutsche Forschungsgemeinschaft (DFG - 469666862, DFG - CR 312/5-2); Italy: Istituto Nazionale di Fisica Nucleare (ICSC, NextGenerationEU), Ministero dell'Università e della Ricerca (NextGenEU 153D23001490006 M4C2.1.1, NextGenEU I53D23000820006 M4C2.1.1, NextGenEU I53D23001490006 M4C2.1.1); Japan: Japan Society for the Promotion of Science (JSPS KAKENHI JP22H01227, JSPS KAKENHI JP22H04944, JSPS KAKENHI JP22KK0227, JSPS KAKENHI JP23KK0245); Norway: Research Council of Norway (RCN-314472); Poland: Ministry of Science and Higher Education (IDUB AGH, POB8, D4 no 9722), Polish National Agency for Academic Exchange (PPN/PPO/2020/1/00002/U/00001), Polish National Science Centre (NCN 2021/42/E/ST2/00350, NCN OPUS 2023/51/B/ST2/02507, NCN OPUS nr 2022/47/B/ST2/03059, NCN UMO-2019/34/E/ST2/00393, UMO-2020/37/B/ST2/01043, UMO-2021/40/C/ST2/00187, UMO-2022/47/O/ST2/00148, UMO-2023/49/B/ST2/04085, UMO-2023/51/B/ST2/00920); Portugal: Foundation for Science and Technology (FCT); Spain: Generalitat Valenciana (Artemisa, FEDER, IDIFEDER/2018/048, ASFAE/2022/008), Ministry of Science and Innovation (MCIN \& NextGenEU PCI2022-135018-2, MICIN \& FEDER PID2021-125273NB, RYC2019-028510-I, RYC2020-030254-I, RYC2021-031273-I, RYC2022-038164-I), Ministerio de Ciencia, Innovación y Universidades/Agencia Estatal de Investigaci\'on (PID2022-142604OB-C22); Sweden: Carl Trygger Foundation (Carl Trygger Foundation CTS 22:2312), Swedish Research Council (Swedish Research Council 2023-04654, VR 2018-00482, VR 2021-03651, VR 2022-03845, VR 2022-04683, VR 2023-03403), Knut and Alice Wallenberg Foundation (KAW 2018.0458, KAW 2019.0447, KAW 2022.0358); Switzerland: Swiss National Science Foundation (SNSF - PCEFP2\_194658); United Kingdom: Leverhulme Trust (Leverhulme Trust RPG-2020-004), Royal Society (NIF-R1-231091); United States of America: U.S. Department of Energy (ECA DE-AC02-76SF00515), Neubauer Family Foundation.

%
%


%
%
%
\printbibliography

\clearpage
\input{atlas_authlist}
%

%
%

%

%
%
%
%

\end{document}

%% file: atlas_authlist.tex
 
\begin{flushleft}
\hypersetup{urlcolor=black}
{\Large The ATLAS Collaboration}

\bigskip

\AtlasOrcid[0000-0002-6665-4934]{G.~Aad}$^\textrm{\scriptsize 105}$,
\AtlasOrcid[0000-0001-7616-1554]{E.~Aakvaag}$^\textrm{\scriptsize 17}$,
\AtlasOrcid[0000-0002-5888-2734]{B.~Abbott}$^\textrm{\scriptsize 124}$,
\AtlasOrcid[0000-0002-0287-5869]{S.~Abdelhameed}$^\textrm{\scriptsize 120a}$,
\AtlasOrcid[0000-0002-1002-1652]{K.~Abeling}$^\textrm{\scriptsize 56}$,
\AtlasOrcid[0000-0001-5763-2760]{N.J.~Abicht}$^\textrm{\scriptsize 50}$,
\AtlasOrcid[0000-0002-8496-9294]{S.H.~Abidi}$^\textrm{\scriptsize 30}$,
\AtlasOrcid[0009-0003-6578-220X]{M.~Aboelela}$^\textrm{\scriptsize 46}$,
\AtlasOrcid[0000-0002-9987-2292]{A.~Aboulhorma}$^\textrm{\scriptsize 36e}$,
\AtlasOrcid[0000-0001-5329-6640]{H.~Abramowicz}$^\textrm{\scriptsize 158}$
% ...
\bigskip
\\

% ...
$^{17}$Department for Physics and Technology, University of Bergen, Bergen; Norway.\\
% ...
$^{30}$Physics Department, Brookhaven National Laboratory, Upton NY; United States of America.\\
% ...
$^{36}$$^{(a)}$Facult\'e des Sciences Ain Chock, Universit\'e Hassan II de Casablanca;$^{(b)}$Facult\'{e} des Sciences, Universit\'{e} Ibn-Tofail, K\'{e}nitra;$^{(c)}$Facult\'e des Sciences Semlalia, Universit\'e Cadi Ayyad, LPHEA-Marrakech;$^{(d)}$LPMR, Facult\'e des Sciences, Universit\'e Mohamed Premier, Oujda;$^{(e)}$Facult\'e des sciences, Universit\'e Mohammed V, Rabat;$^{(f)}$Institute of Applied Physics, Mohammed VI Polytechnic University, Ben Guerir; Morocco.\\
% ...
$^{46}$Physics Department, Southern Methodist University, Dallas TX; United States of America.\\
% ...
$^{50}$Fakult\"{a}t Physik, Technische Universit{\"a}t Dortmund, Dortmund; Germany.\\
% ...
$^{56}$II. Physikalisches Institut, Georg-August-Universit\"{a}t G\"ottingen, G\"ottingen; Germany.\\
% ...
$^{105}$CPPM, Aix-Marseille Universit\'e, CNRS/IN2P3, Marseille; France.\\
% ...
$^{120}$$^{(a)}$New York University Abu Dhabi, Abu Dhabi;$^{(b)}$United Arab Emirates University, Al Ain; United Arab Emirates.\\
% ...
$^{124}$Homer L. Dodge Department of Physics and Astronomy, University of Oklahoma, Norman OK; United States of America.\\
% ...
$^{158}$Raymond and Beverly Sackler School of Physics and Astronomy, Tel Aviv University, Tel Aviv; Israel.\\
% ...
$^{a}$ Also Affiliated with an institute covered by a cooperation agreement with CERN.\\
$^{b}$ Also at An-Najah National University, Nablus; Palestine.\\
$^{c}$ Also at Borough of Manhattan Community College, City University of New York, New York NY; United States of America.\\
$^{d}$ Also at Center for High Energy Physics, Peking University; China.\\
$^{e}$ Also at Center for Interdisciplinary Research and Innovation (CIRI-AUTH), Thessaloniki; Greece.\\
$^{f}$ Also at CERN, Geneva; Switzerland.\\
% ...

\end{flushleft}
